\documentclass[twocolumn,preprintnumbers,amsmath,amssymb]{revtex4}

\usepackage{graphicx}
\usepackage{bm}

\newcommand{\beq}{\begin{equation}}
\newcommand{\eeq}{\end{equation}}
\newcommand{\bea}{\begin{eqnarray}}
\newcommand{\eea}{\end{eqnarray}}
\newcommand{\re}{\text{Re}}

\let\a=\alpha   \let\g=\gamma  \let\d=\delta 
\let\z=\zeta       
                 
 \let\t=\tau    
   \let\o=\omega

\let\io=\infty

\begin{document}

\title
{Mode-locking transitions in nano-structured weakly disordered lasers}

\author{L. Angelani$^1$, C. Conti$^{2,3}$, L. Prignano$^{4}$, G. Ruocco$^{3,4}$, and F. Zamponi$^{5}$}

\affiliation{
$^1$Research center SMC INFM-CNR, c/o Universit\`a
di Roma ``La Sapienza,'' I-00185, Roma, Italy \\
$^2$Centro studi e ricerche ``Enrico Fermi'', Via Panisperna 89/A,
I-00184, Roma, Italy   \\
$^3$Research center Soft INFM-CNR, c/o Universit\`a di
Roma ``La Sapienza,'' I-00185, Roma, Italy \\
$^4$Dipartimento di Fisica, Universit\`a di Roma ``La Sapienza,''
I-00185, Roma, Italy \\
$^5$Service de Physique Th\'eorique, 
CEA Saclay, 91191 Gif sur Yvette Cedex, France
}

\begin{abstract}
We report on a statistical approach to mode-locking
transitions of nano-structured laser cavities characterized by an
enhanced density of states. We show that the equations for the interacting
modes can be mapped onto a statistical model exhibiting a first order 
thermodynamic transition, with the average mode-energy playing the role of inverse temperature.
The transition corresponds to a phase-locking of modes.
Extended modes lead to a mean-field like model, while in presence of
localized modes, as due to a small disorder, the model has short range interactions.
We show that simple scaling arguments lead to observable differences between transitions involving extended modes
and those involving localized modes. We also show that the dynamics of the
light modes can be exactly solved, predicting a jump in the relaxation time
of the coherence functions at the transition.
Finally, we link the
thermodynamic transition to a topological singularity of the phase space,
as previously reported for similar models.
\end{abstract}

\maketitle

\section{Introduction}
Laser mode-locking (ML) is well known in standard optical
resonators, which are characterized by equi-spaced resonances
\cite{YarivBook,SiegmanBook}. ML in such a kind of systems is a
valuable route for the generation of ultra-short pulses, in
particular when it is ``self-starting'', as due to the nonlinear
interaction between laser longitudinal modes \cite{Haus00}. Given
the growing interest in high-Q microresonators and photonic
crystals \cite{JoannopoulosBook,SakodaBook}, it is interesting to
consider ML in integrated devices, which could trigger a new
generation of highly-miniaturized lasers emitting ultra-short
pulses (see e.g. \cite{Liu06}).

In this respect, there is a remarkable difference between standard
resonators and nano-structured cavities; indeed the latter are
characterized by a non-uniform distribution of resonances, given
by a strongly modulated density of states (DOS) \cite{SakodaBook}.
This situation favors a new formulation of the analysis of the
self-mode locking transition, based on a mean-field thermodynamic
approach: this is the topic of the present paper, also
including the effect of some disorder in the system. The
thermodynamic approach to multi-mode interactions in various
physical frameworks is well established \cite{HakenBook}. For
example,  it was recently applied to transverse-mode interaction
in resonators \cite{Papoff04,Cabrera06}, as well as to
standard-laser mode-locking transition \cite{Gordon02,Gordon03,Sierks98}.
This transition can be described in terms of an effective
temperature $T$, which encompasses the level of noise due to
spontaneous emission and the amount of energy stored into each
mode. At high $T$ the mode-phases are independent and rapidly
varying (``free-run'' or ``paramagnetic phase''); conversely, either reducing the
spontaneous emission noise or increasing the pumping rate, a
low-temperature (``ferromagnetic'') phase can be reached,
corresponding to the mode-phases locked at the same value.

A paradigm that has been recently shown to be very effective for
describing the nonlinear interaction of many ``modes'', and the
resulting phase transitions and/or kinetic arrest is the potential
energy landscape (PEL) approach (see e.g.
\cite{Debenedetti01,WalesBook}).
The PEL, as a manifold in the configurational phase space,
has many stationary points (typically minima and saddles)
\cite{angelani00}, whose distribution strongly affects the thermodynamics
(and the dynamics) of the system. Recently this
paradigm, developed to investigate the glass transition phenomena,
has been applied to the field of photonics, including optical
solitons \cite{Conti05,Conti06} and random-lasers
\cite{Angelani06, Angelani06b}. In this respect, it is worth to
note that the geometrical interpretation of the laser threshold
was recognized since the beginning of laser theory, and is
considered as one of the successful applications of
catastrophe-theory, which classifies the singularities of
multi-dimensional manifolds \cite{HakenBook,GilmoreBook}. It is
not surprising, therefore, that the mode locking transition can be
interpreted according to the thermodynamic/topological transition
point of view. Extending the topological approach to the
nano-laser is interesting for different reasons: on one hand this
provides an elegant and comprehensive theoretical framework to
laser theory, on the other hand it can be relevant from a
fundamental physical perspective. Indeed, in recent literature the
link between geometry and thermodynamics has been strongly
debated \cite{Casetti00,Franzosi04,Angelani03,Zamponi03}, and it is still to be
established if this theoretical circumstance has physical
consequences. Thus it is important to identify physical systems
which can be treated by analytical solvable models (for what
concerns both the thermodynamical and the topological properties).

Here we show that mode-locked laser nano-cavities fall within this
category. In addition we also report on explicit expressions of the first-order
coherence function of the laser emission. This analysis predicts a jump in the
relaxation time (and correspondingly in the laser line-width) at the mode-locking transition.
The scaling properties of the threshold average mode-energy at the transition are found to be strongly
sensible to the degree of localization of the involved modes.

The outline of this manuscript is as follows: in section
\ref{sec1} we will recall the mode-coupling approach to multi-mode
lasers; in section \ref{secLoc} we will discuss the physical
signatures of transitions involving either localized or delocalized
modes; in section \ref{secThe} we will report on the thermodynamic
approach; the analysis of the topological origin of the laser
transition is given in \ref{secTop}; in section \ref{secDyn} we briefly discuss
the solution of the dynamics of the model; conclusions are drawn in
section \ref{secCon}.

\section{Multi-mode laser equations}
\label{sec1}
The coupled mode theory equations in a nonlinear cavity 
can be written in the form \cite{Gordon03,Angelani06b,Lamb64,Bryan73}
\begin{equation}\begin{split}
\frac{da_s}{dt} &=-\frac{\partial{H_I}}{\partial{a_s^*}}-\alpha_s a_s(t) 
+ (\g_s- g_s |a_s(t)|^2) a_s(t) + \eta_s(t) \\
&=-\frac{\partial\mathcal{H}}{\partial{a_s^*}}+\eta_s(t) \ ,
\label{finale}\end{split}
\end{equation}
with
\begin{equation}
\mathcal{H}=H_o+H_I \ ,
\end{equation}
and
\begin{equation}\begin{split}
H_o &= \sum_s  ( \alpha_s - \gamma_s) |a_s|^2 + \frac12 g_s |a_s|^4 \\
&= \sum_s V_s(a_s)
\text{.}
\end{split}\end{equation}
In (\ref{finale}) $s=1,2,..,N$ with $N$ the total number of modes, while $a_s$ is
the complex amplitude of the mode at $\omega_s$, such that $\mathcal{E}_s=\omega_s |a_s|^2$
is the energy stored in the mode.
Radiation losses and material absorption,
are represented by the coefficient $\alpha_s$, while $\g_s - g_s |a_s|^2$ represents the 
saturable gain term
and, as usual, the quantum noise term due to spontaneous emission is given by a random term
$\eta_s$ such that
$\langle \eta_s(t)\eta_p(t') \rangle=2k_B T_{bath} \delta_{sp}\delta(t-t')$,
where $k_B$ is the Boltzmann constant and $T_{bath}$ is an effective temperature \cite{Angelani06b}.
The nonlinear interaction term is
\begin{equation}
H_I=\re [\frac{1}{4}\sum_{\{spqr\}}g_{spqr}a_s a_p a_q^* a_r^*],
\end{equation}
where the sum is extended over all mode resonances such that $\omega_s+\omega_p=\omega_q+\omega_r$.
The term $s=p=q=r$ is not included as it is already described by $g_s$; 
the Hamiltonian $H_I$ describes mode interaction 
due to the nonlinearity of the gain medium.
The field overlap is given by
\begin{equation}
\begin{split}
g_{spqr} = &\displaystyle\frac{\sqrt{\omega_s \omega_p \omega_q \omega_r}}{2i} \int_V
\chi_{\alpha \beta \gamma \delta}(\omega_s;\omega_q,\omega_r,-\omega_p,\textbf{r})\times \\
& \times {E}_s^{\alpha}{E}_p^{\beta}{E}_q^{\gamma}{E}_r^{\delta}dV \ .
\end{split}
\label{gs}
\end{equation}
where $V$ is the cavity volume, $\chi$ is the third order-susceptibility tensor due to the resonant medium and $E_p^\alpha$ are
the components  ($\alpha=1,2,3$) of the vectorial mode of the cavity at the
resonance $\omega_p$.
$\chi$ is given, in the simplest formulation, by the Lamb theory \cite{Lamb64,Bryan73} and,
neglecting mechanisms like 
self- and cross-phase modulation (which give phase-independent contribution to the 
relevant Hamiltonian, see below)
can be taken as real-valued;
under standard approximations, the tensor $g$ is a quantity symmetric with respect to the exchange of 
any couple of indexes.

By letting $a_s(t)=A_s(t)\exp\left[i\varphi_s(t)\right]$,
the $\mathcal{H}$ can be rewritten as
\begin{equation}
\mathcal{H}(G,\varphi)=H_o+\sum_{\{spqr\}}G_{spqr}\cos(\varphi_s+\varphi_p-\varphi_q-\varphi_r)
\label{Hcal}
\end{equation}
where $H_o=\sum_s V_s(A_s)$ only depends on the amplitudes
and $G_{spqr}=2g_{spqr}A_sA_pA_qA_r$.
As discussed in the literature \cite{Gordon03} Eqs. (\ref{finale})
are Langevin equations for a system of $N$ particles moving in $2N$
dimensions and the invariant measure is given by $\exp(-\mathcal{H}/k_B T_{bath})$.

In a standard laser the resonant frequencies are equispaced and this gives rise
to various formulations of laser thermodynamics, which are based on the fact that
the  $\omega_s+\omega_p=\omega_q+\omega_r$ will only involve a limited number of interacting modes
 \cite{Gordon02,Gordon03}.
The situation is drastically different for nano-structured systems displaying a photonic-band gap.
It is indeed well established that in proximity of the  band-edge,
a DOS enhancement with respect to vacuum is obtained. All the corresponding modes will have overlapping resonance such that $\omega_s\cong\omega_0$,
(where $\omega_0$ is the position of the peak in the density of states, which is assumed
to be in correspondence of the resonance of the amplifying atomic medium);
additionally the resonance condition $\omega_s+\omega_p=\omega_q+\omega_r$
need not to be exactly satisfied, but it sufficient that linewidth of the corresponding modes needs
to be overlapped for a relevant interaction \cite{MeystreBook}.
Hence for such a kind of system it is interesting to consider
a mean field regime where all the modes interact in a limited spectral region around $\omega_0$.
For the mode-locking transition one can limit to consider the phase dynamics.
Indeed ML entails the passage from a regime in which the
mode-phases are independent and rapidly varying
(``free-run'' regime or ``paramagnetic phase'' in the following) \cite{Lamb64}
on times scales of the order of $10$ fs \cite{Brunner83},
to a regime in which they are all locked at the same values (``ferromagnetic phase'').
In correspondence of this
transition the laser output switches from a continuous wave noisy emission to an highly modulated signal
(which is a regular train of short-pulses for equi-spaced resonances).
Mode-amplitude dynamics is not affected (at the first
approximation) by the onset of ML. Indeed for lasing modes $\g_s > \a_s$, so that the potential
$V_s(A_s)$ has a single minimum in $A_s = \sqrt{(\g_s-\a_s)/g_s}$. Thus the amplitudes of the
lasing modes will fluctuate around this minimum: we neglect these fluctuations, which are small
if the potential well is deep enough, and
threat the $A_s$ as {\it quenched} variables
($\mathcal{E}_s=\omega_s |a_s|^2\cong \langle \mathcal{E}_s\rangle\equiv \omega_0 A^2$
for the relevant modes).
The relevant subspace spanned by the system is given by the phases, that are taken as the
dynamic variables (see, for example, \cite{YarivBook} for a discussion of the role
of mode-phases with respect to amplitudes in ML processes).
\section{Localized versus delocalized modes in the self-mode-locking transition}
\label{secLoc}
In previous works \cite{Angelani06,Angelani06b} we made reference to a completely random resonator,
for which the $G$ coefficients were taken Gaussian distributed with zero mean.
This is the natural approach when dealing with strongly disordered resonators,
 in which the involved modes can be localized or
delocalized in the structure and the corresponding resonances and spatial distribution can have different degrees
of overlaps (as in \cite{Mujumdar04}).
Here we make reference to the opposite regime, corresponding to case in which
the structure is quasi-ordered, with the presence of a small amount of disorder.
The disorder is such that the variations in the coupling coefficients $g_{spqr}$ can be taken as negligible
with respect to their statistical average $\langle g_{spqr}\rangle \simeq g$,
however it is sufficient to induce
the existence of a tail of localized modes in the photonic band gap \cite{John87}.

As discussed above, we consider mode-resonances packed in a small
spectral region $\Delta \omega$ ( if compared with the central carrier angular region, i.e.
$\Delta \omega<<\omega_0$).
This kind of system is very different from the standard laser cavity, with
equispaced mode-frequencies.
A prototypical structure is given by a photonic crystal doped by active materials.
In the absence of disorder the involved modes are Bloch modes,
which are extended over the whole sample and are absent in the forbidden band gap. Their
DOS is peaked at the band-gap edge \cite{SakodaBook}. In this case, the modes have overlapping resonances
(the width of each spectral line being determined by material and radiation losses)
and they also have not-negligible spatial overlap
(with exception of those mode-combinations which are vanishing for symmetry reasons).
In the presence of a small amount of disorder it is well established \cite{John87} that a tail
of localized states appears in the photonic band gap. Hence the localized states also have
overlapping resonances in tiny spectral regions in proximity of the band-edge. Their
spatial overlap can be strongly reduced with respect to the Bloch modes, however
the exponential tails of their spatial profiles are expected to provide not vanishing values for $g$.
Localized states can also be introduced intentionally, e.g. by using defects in a planar PC slab-waveguide \cite{JoannopoulosBook},
or in coupled cavity systems (see e.g. \cite{Liu06} and references therein).
In this case the spatial overlap and the resonance frequencies can be tailored at will.
Thus, in the general case, extended and localized states can be involved in the mode-locking
transitions here considered. However, simple scaling arguments lead to the conclusion that
the two kind of modes display a macroscopic difference in their ``thermodynamics''.

{\it Extended modes}.
We start considering  the extended modes.
From the normalization, the modules of the eigenvectors $E_s$ are such that $\int dV |E_s|^2 = const.$
Denoting with $V_o$ the volume over which a given mode is different from zero, one has $E_s \sim V_o^{-1/2}$.
For extended modes $V_o \propto V$;  in addition, most of the mode-overlaps are not vanishing,
hence the sum in (\ref{Hcal}) will involve all the modes (within the spectral region $\Delta\omega$),
and this implies that $H_I\propto V^4\propto N^4$.
The Hamiltonian is hence more than extensive (for an extensive one $H \propto N$); 
however a thermodynamics approach is still possible if one accepts that the effective
temperature will depend on the volume: the effective temperature will be taken
as proportional to $N^{-2}\propto V^{-2}$, as detailed below, so that the resulting effective Hamiltonian will be proportional to $V$.
Physically this corresponds to the fact that the energy (and hence the pumping rate) needed to
induced the ML transition will grow with the number of modes, if only extended modes are involved.
For the overlap coefficients, in the absence of strong disorder $g_{spqr}\cong g\neq 0$, and
the coupling $g$ scales with the inverse volume, $g \propto V^{-1} \propto N^{-1}$ ($|E|^4 \sim V^{-2}$ and $\int dV \chi \sim V$).
Thus considering the invariant measure
($A_s\simeq A$, $g_{spqr}\simeq g$ e $G_{spqr}\simeq G=g\,A^4$), one has
\begin{equation}
\begin{array}{l}
\exp(-\displaystyle\frac{H_I}{k_B T_{bath}})=\\
\exp[-\displaystyle\frac{g A^4} {k_B T_{bath}}\sum_{spqr} \cos(\varphi_s+\varphi_p-\varphi_q-\varphi_r)]=\\
\exp[-\displaystyle\frac{g A^4 N^3} {k_B T_{bath}} \frac{1}{N^3} \sum_{spqr} \cos(\varphi_s+\varphi_p-\varphi_q-\varphi_r)]\equiv\\
\exp[-\beta H^{(ext)}] \ ,
\end{array}
\end{equation}
where $\beta\equiv |g| A^4 N^3/ k_B T_{bath}\equiv 1/T\propto N^2$ is an inverse adimensional temperature, and the mode-phase dependent
(extensive) Hamiltonian is given by (within an irrelevant additive term)
\begin{equation}
H^{(ext)}=\frac{1}{ N^3}\sum_{spqr}\left[1-\cos(\varphi_{s}+\varphi_{p}-\varphi_{q}-\varphi_{r})\right] \ .
\label{Hfin}
\end{equation}
In (\ref{Hfin}) we have used the fact that $g<0$ in all the physically relevant regimes.
No transition is expected for $g>0$, as detailed below.
Note that, as we assumed that the condition $\o_s + \o_p = \o_q + \o_r$ is not
exactly satisfied, and possibly due to the presence of disorder, the integrand in (\ref{gs})
might have oscillations or fluctuations in sign that can affect the scaling of $g$ with volume (the case 
of completely random fluctuations lead to a scaling $g \propto V^{-3/2}$ and to an extensive
Hamiltonian as in \cite{Angelani06,Angelani06b}). Intermediate regimes might be present depending
on the strength of the disorder.

{\it Localized modes}.
Next we consider localized modes, that exponentially decay in space.
Defining $V_o$ as above, it turns that it does not scale with the volume $V$ of the sample,
but can be written as $V_o=L_o^3$ where $L_o$ is an average localization length.
The overlap coefficients $g_{spqr}$ will have a statistics strongly peaked around some average value
$g$, measuring the average amount of spatial overlap between localized modes that are neighborhood in space.
Hence the sum in the Hamiltonian will only involve first neighborhoods and $H_I\propto V$, while
$g$ will not depend on the size of the system.
For the invariant measure we will have (the angular bracket in $\langle spqr \rangle$ denoting sum over first neighborhoods):
\begin{equation}
\begin{array}{l}
\exp(-\displaystyle\frac{H_I}{k_B T_{bath}})=\\
\exp[-\displaystyle\frac{g A^4} {k_B T_{bath}}\sum_{\langle spqr \rangle} \cos(\varphi_s+\varphi_p-\varphi_q-\varphi_r)]=\\
\exp[-\beta H^{(loc)}] \ ,
\end{array}
\end{equation}
where $\beta\equiv |g| A^4 / k_B T_{bath}\equiv 1/T$
will not depend on the system size (we stress that the system size must be such that a
large number of localized modes are present), and the Hamiltonian (within irrelevant additive constants) is
\begin{equation}
H^{(loc)}= \sum_{\langle spqr \rangle} [1-\cos(\varphi_s+\varphi_p-\varphi_q-\varphi_r)] \ .
\label{Hloc}
\end{equation}

\noindent Summarizing, {\it if the transition involves extended modes, the effective temperature
for the critical transition is expected to depend on the size of the system; 
conversely for localized modes the critical temperature
will be independent on the system size}.

Unfortunately, at variance with fully-connected (or ``mean field'') models as (\ref{Hfin}),
analytical treatment of short-range Hamiltonians as (\ref{Hloc})
is almost always impossible and
the analysis can only be numerically performed.
However, it is well established 
within the statistical physics community that mean-field
models obtained from first-neighborhood systems conserve most of the thermodynamics properties,
and more specifically the existence of a thermodynamic transition,
at least above the so called ``lower critical dimension'' $d_l$
(for example, for the Ising model is $d_l=1$, for the $XY$ model is $d_l=2$).
Our model falls in the class of $XY$ models so we expect that the transition
exists, as long as $d>2$, also in the case of
localized modes, and the following analysis applies at least qualitatively.
Since (\ref{Hfin}) can be analytically treated
for thermodynamic, topological and dynamic properties we will limit to this model in the following.
The existence of a thermodynamic/topological transition is expected in the general case, while the different
scaling properties of the effective temperature enables to discern localized and delocalized interactions.
\section{Thermodynamics}
\label{secThe}
In this Section we study the thermodynamics of the laser Hamiltonian in the mean field approximation
(\ref{Hfin}),
within the quenched amplitudes approximation.
The partition function is
\begin{equation}
Z=\int d\varphi \ e^{-\beta H(\varphi)}
\end{equation}
where $H$ is
\begin{equation}
H=\frac{1}{ N^3}\sum_{spqr}\left[1-\cos(\varphi_{s}+\varphi_{p}-\varphi_{q}-\varphi_{r})\right] \ .
\label{2x2x*}
\end{equation}

The Hamiltonian (\ref{2x2x*}) is very similar to that defining the $k$-trigonometric model ($k$-TM) for $k=4$,
introduced in \cite{Angelani03} with the aim of studying the relation between phase transitions
and topological property of the potential energy surface \cite{Casetti00,Angelani05,ktm_JCP}.\\
Defining  the ``magnetization''
\begin{equation}
z=\frac{1}{N}\sum_ie^{i \varphi_i}=\xi e^{i \psi},
\label{compmagn}
\end{equation}
where $\xi$ and $\psi$ depend on $\{\varphi_i\}$, we have
\begin{eqnarray}
H & = &{\textrm{Re}}\left[\frac{1}{N^3}\sum_{spqr}\left[1-\exp i(\varphi_s+\varphi_p-\varphi_q-\varphi_r)\right]\right]=
\nonumber \\
& = &\textrm{Re}\left[\frac{1}{N^3}N^4(1-z^2z^{*2})\right]=N (1-\xi^4) \ .
\label{Ham}
\end{eqnarray}
By definition a vanishing $z$ denotes uncorrelated phase, as in the ``free run'' regime, conversely
if $z\neq0$ the phase of the modes are correlated and locked.
The thermodynamics of the mean-field model is exactly solved by neglecting the correlations between different degrees of freedom,
and obtaining an effective Hamiltonian that contains a parameter to be determined self-consistently.
Introducing the mean (complex) ``magnetization'' $\zeta=\langle \textrm{e}^{i\varphi}\rangle$, and substituting in equation (\ref{2x2x*}) the expression
\begin{eqnarray*}
e^{j(\varphi_s+\varphi_p-\varphi_q-\varphi_r)}\to e^{j\varphi_s}\langle e^{j\varphi_{p}}\rangle\langle e^{-j\varphi_{q}}\rangle \langle e^{-j\varphi_{r}}\rangle +\\
+\langle e^{j\varphi_{s}}\rangle e^{j\varphi_{p}}\langle e^{-j\varphi_{q}}\rangle \langle e^{-j\varphi_{r}}\rangle+ \\
+\langle e^{j\varphi_{s}}\rangle \langle e^{j\varphi_{p}}\rangle e^{-j\varphi_{q}}\langle e^{-j\varphi_{r}}\rangle+ \\
+\langle e^{j\varphi_{s}}\rangle \langle e^{j\varphi_{p}}\rangle \langle e^{-j\varphi_{q}}\rangle e^{-j\varphi_{r}}- \\
-3\langle e^{j\varphi_{s}}\rangle\langle e^{j\varphi_{p}}\rangle\langle e^{-j\varphi_{q}}\rangle\langle e^{-j\varphi_{r}}\rangle=
\end{eqnarray*}
\begin{equation}
= 2\zeta\zeta^*(e^{j\varphi}\zeta^*+e^{-j\varphi}\zeta)-3\zeta^2\zeta^{*2}=4\zeta^3\cos\varphi-3\zeta^4
\end{equation}
where the last equality stands because we have chosen $\zeta$ to be real without loss of generality
(corresponding to choosing a particular magnetization of the low temperature state),
the effective Hamiltonian $h$ per degree of freedom  reads as
\begin{equation}
h(\varphi)=
1+3\zeta^4-4\zeta^3\cos\varphi \ .
\label{h}
\end{equation}
The self-consistent equation for $\zeta$ turns out to be
\begin{equation}
\zeta=\langle\cos\varphi\rangle_h=\frac{I_1(4 \beta \zeta^3)}{I_0(4 \beta \zeta^3)} \ ,
\label{xi}
\end{equation}
where $I_0(\alpha)=(2\pi)^{-1}\int_0^{2\pi} d\varphi  \exp(\alpha \cos\varphi)$
and $I_1(\alpha)=I_0'(\alpha)$ are the modified Bessel function of order 0 and 1,
and $\langle \dots \rangle_h$ is the average over the probability distribution
\begin{equation}
P(\varphi) = \frac{e^{-\beta h(\varphi)}}{\cal Z} \ ,
\end{equation}
with
\begin{equation}
{\cal Z} = \int_{0}^{2\pi} d\varphi \ e^{-\beta h(\varphi)} = 2\pi \ e^{-\beta (1+3\zeta^4)}\ I_0(4 \beta  \zeta^3) \ .
\end{equation}
The solution of Eq. (\ref{xi}) are the extrema of the free energy $f$ as a function of $\zeta$
\begin{equation}
\beta f = - \ln{\cal Z} = \beta(1+3\zeta^4) - \ln{2\pi I_0(4 \beta \zeta^3)} \ ,
\end{equation}
whose absolute minimum is the thermodynamical stable solution.
\begin{figure}[t]
\begin{center}
\includegraphics[width=.47\textwidth]{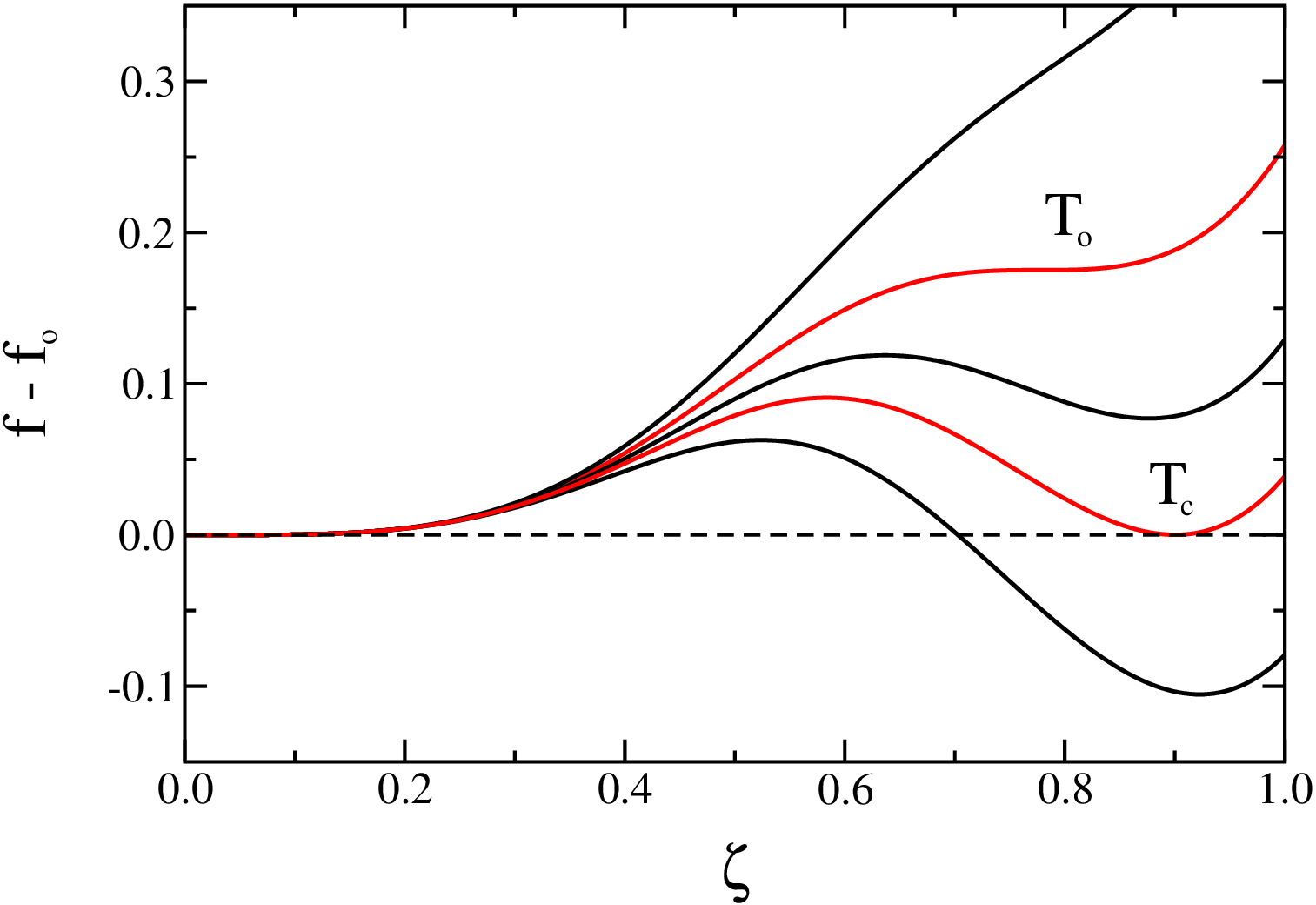}
\end{center}
\caption{(Color online)
Free energy $f(\zeta)-f(\zeta\!=\!0)$ as a function of magnetization $\zeta$ at different temperatures.
From high to low: $T=0.910$, $T_o=0.717$, $T=0.616$, $T_c = 0.548$, $T=0.504$.
$T_o$ marks the appearance of the unstable minimum, $T_c$ is the transition
temperature at which the solution with $\zeta >0$ becomes thermodynamically stable.
\label{fig1}}
\end{figure}

\begin{figure}[t]
\begin{center}
\includegraphics[width=.45\textwidth]{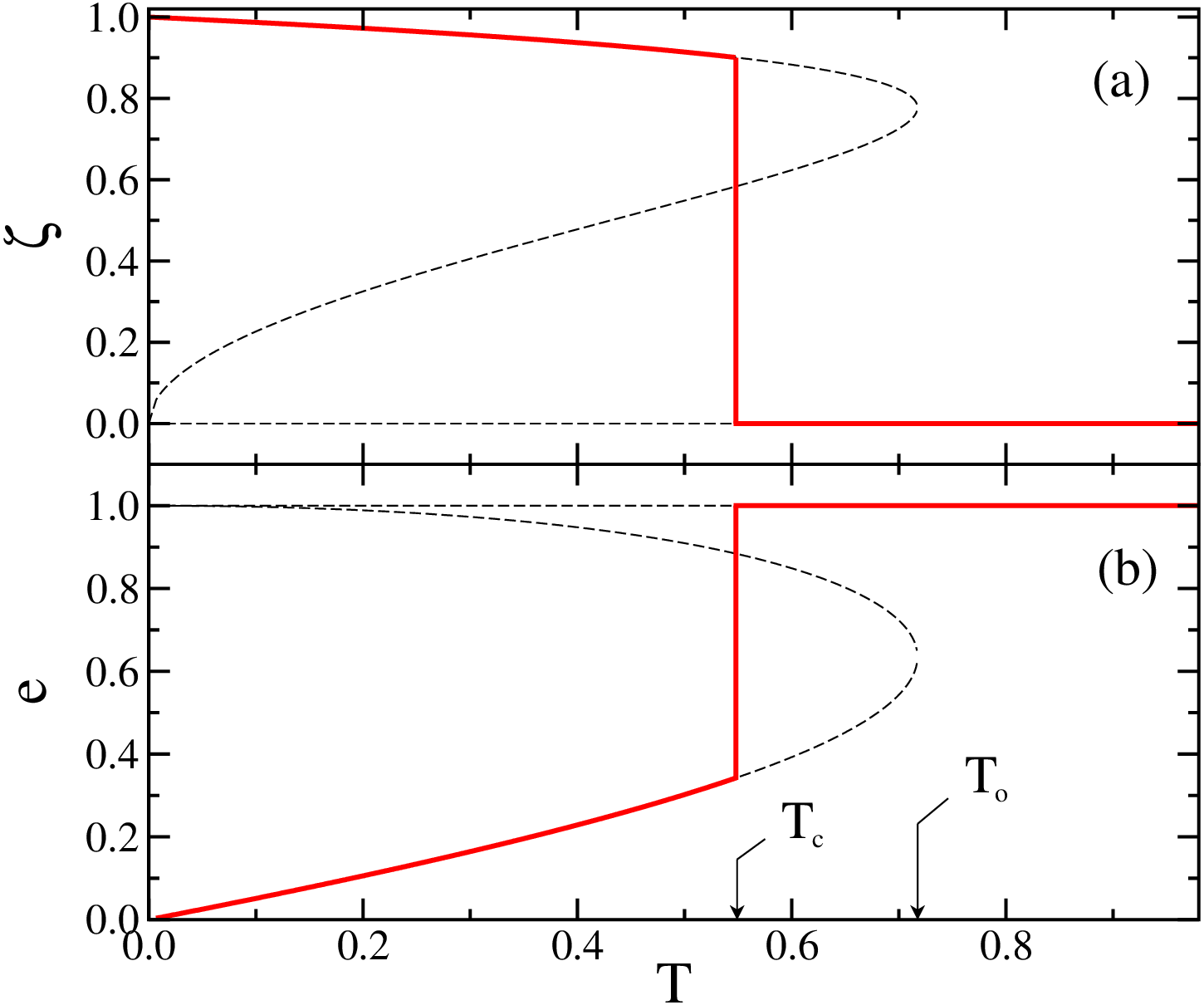}
\end{center}
\caption{(Color online)
Magnetization $\zeta$ - part (a) - and energy $e$ - part (b) -
as a function of temperature. Full lines correspond to the thermodynamically stable solution of
Eq. (\ref{xi}), dashed lines are the unstable solutions.
At $T_c=0.548$ a first-order thermodynamic phase transition
takes place, while at $T_o=0.717$ unstable solutions appear.
\label{fig2}}
\end{figure}

The value $\zeta=0$, corresponding to the paramagnetic solution, always solves equation (\ref{xi}),
but it gives the stable (lower free-energy) solution only for low $\beta$ (high $T$).
On lowering $T$, at $T_o=0.717$
other solutions appear, such that $\zeta\neq 0$.
However, the stable solution is still the paramagnetic one $\zeta=0$.
At $T_c=0.548$ the solution
$\zeta \neq 0$ becomes the stable one, and a first-order phase transition takes place.
In Fig. \ref{fig1} the $\zeta$-dependence of the free energy $f$
is reported for different temperatures.
The stable solution $\zeta(T)$ is shown in Fig. \ref{fig2}a (full line) while dashed lines denote
unstable solutions (local minimum and maximum).
In  Fig. \ref{fig2}b the $T$-dependence of the energy
\begin{equation}
e=-\frac{\partial}{\partial \beta} \ln{\cal Z} = 1-\zeta^4 \ ,
\label{hmean}
\end{equation}
is shown for the stable (full line) and unstable (dashed lines) solutions.\\
A remark on the sign of the coupling $g$. For $g\!>\!0$ the sign of the $\cos$ term in the Hamiltonian
(\ref{2x2x*}) is positive and the self-consistent equation reads
$\zeta  =-I_1(\beta\Delta4\zeta^3) / I_0(\beta\Delta4\zeta^3) $, which has the only solution
$\zeta = 0$.
Then, in this case the phase transition does not take place.

\section{Topology}
\label{secTop}
After having ascertained the existence of a first-order phase-transition,  we consider the property
of the stationary points (saddles) of the  potential energy landscape of the system \cite{Angelani03}.
As said in the Introduction, in recent works \cite{Casetti00,Franzosi04}, it has been conjectured
that phase transitions are signaled by discontinuities in the configuration space topology.
More precisely, for  a system defined by a continuous potential energy
function $V(q)$ ($q$ denotes the N-dimensional vector of the generalized coordinates)
a thermodynamic phase transition occurring at $T_c$  (corresponding to energy $V_c$)
is the manifestation of a topological discontinuity taking place at $V_c$
(``topological hypothesis'').
The most striking consequence of this hypothesis is that the signature of a phase transition
is present in the topology of the configuration space independently on the statistical measure defined on it.
Through Morse theory, topological changes are related to the presence of stationary points of $V$,
and, more specifically, to the discontinuous behavior of invariant quantity defined on them,
as the Euler characteristic $\chi$.
Subsequent works \cite{phi4,topo_1d} have shown that, at least for some model system,
a ``{\it weak topological hypothesis}'' applies in place of the ``strong'' one:
the $V_\theta$ at which a topological transition takes place  does not coincide with the thermodynamic
one $V_c \neq V_{\theta}$, but is related to it by a saddle-map $M$, from equilibrium energy level
to stationary point energy: $M(V_c)=V_{\theta}$.
Then, the role of saddles has been demonstrated to be  of high relevance
for the topological interpretation of thermodynamic transitions.
Here we report on the saddle properties of the considered nano-laser model.

The stationary point $\overline{\varphi}$ are defined by the condition d$H(\overline{\varphi})=0$
and their order is defined as the number of negative eigenvalues of the Hessian matrix $H_{ij}=(\partial{^2H} / \partial{\varphi_i}\partial{\varphi_j})$$|_{\overline{\varphi}}$. To determine the location of stationary point we have to solve the system
\begin{equation}
\frac{\partial H}{\partial\varphi_k}=
4 \xi^3\sin(\varphi_k-\psi)=0,\ \ \ \ \ \   \forall k
\label{staz}
\end{equation}
where we have used Eqs. (\ref{compmagn}) and (\ref{Ham}).\\
A first group of solutions arises for $\xi =0$;
from equation (\ref{Ham}) we have $H=N\big(1-\xi^4\big)$,
and then the stationary points with $\xi(\overline{\varphi})=0$ are located at the energy
$\textrm{e}= H(\overline{\varphi})/N=1$.
Now we restrict ourselves to the region $e\neq 1$ because,
as we will see at the end, the quantities in which we are interested are singular
when $e=1$.
For $e\neq 1$, equation (\ref{staz}) becomes
\begin{equation}
\sin(\varphi_k-\psi)=0\ \ \ \ \ \  \forall k 
\end{equation}
and its solutions are
\begin{equation}
\varphi_k=\left[\psi+ m_k\pi\right]_{{\rm mod}\ 2\pi} 
\label{fij}
\end{equation}
where $m_k=\{0,1\}$.
The unknown constant $\psi$ is found by substituting Eq. (\ref{fij}) in the self-consistency equation
\begin{equation}
\begin{array}{l}
z=\xi e^{i\psi}=N^{-1}\sum_ie^{i\varphi_i}=N^{-1}\sum_ie^{i(\psi+m_i\pi)}=\\N^{-1}e^{i\psi}\sum_i(-1)^{m_i}\ .
\end{array}
\label{zeta}
\end{equation}
Introducing the quantity $n(\overline{\varphi})$ defined by
\begin{equation}
n=N^{-1}\sum_im_i\ ,\ \ \ \ 1-2n= N^{-1}\sum_i(-1)^{m_i} 
\label{ndef}
\end{equation}
we have, from equation (\ref{zeta}),
\begin{equation}
\label{xin}
\xi=1-2n \ .
\end{equation}
As $\xi$ is positive defined, the only solutions are for $n<1/2$:
there are not stationary points with $n>1/2$.
\begin{figure}[t]
\begin{center}
\includegraphics[width=0.5\textwidth]{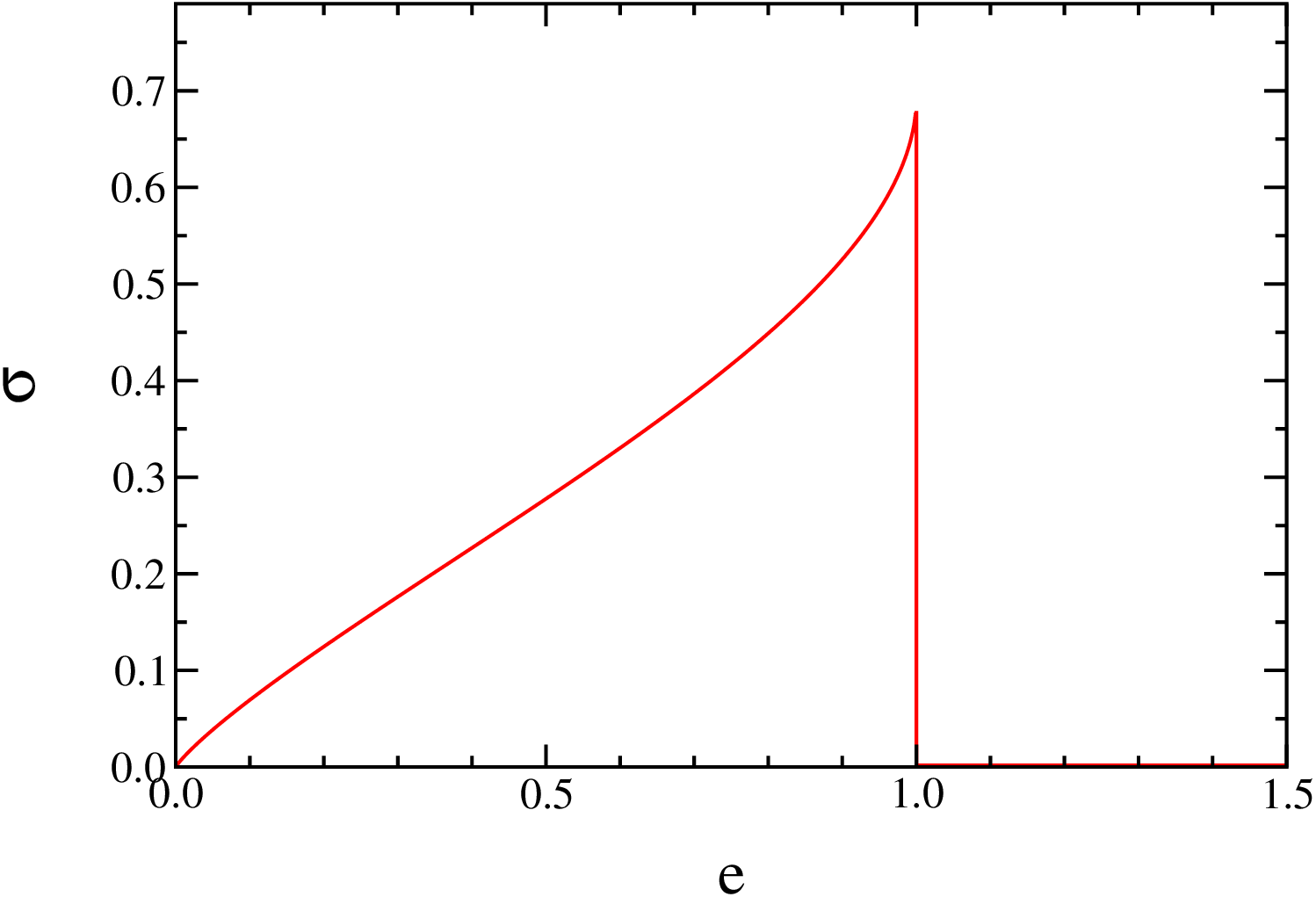}
\end{center}
\caption{(Color online)
The entropy of saddles $\sigma$ as a function of potential energy $e$.}
\label{fig3}
\end{figure}For $n<1/2$
$\psi$ can assume all the values in $[0,2\pi)$ for any choice of the set $\{m_k\}$
and all the stationary points of energy $e\neq 1$ have the form
\begin{equation}
\overline{\varphi}^\textbf{m}_k=\left[\psi + m_k\pi\right]_{{\rm mod} \ 2\pi},\ \textrm{where}\ \textbf{m}=\{m_k\},\ m_k=(0,1) 
\label{stazp}
\end{equation}
under the condition
\begin{equation*}
n=N^{-1}\sum_k m_k<1/2\ .
\end{equation*}
The Hessian matrix is given by
\begin{equation}
\begin{array}{l}
H_{ij} = 4 \xi^2\Big\{-\frac{4}{N}\xi^2\sin(\varphi_i-\psi)\sin(\varphi_j-\psi) -\\
\frac{1}{N}\cos(\varphi_i-\varphi_j)+\delta_{ij}\xi\cos(\varphi_i-\psi)\Big\}\ .
\end{array}
\end{equation}
In the thermodynamic limit it becomes diagonal,
\begin{equation}
H_{ij} \simeq 4\xi^3\delta_{ij}\cos(\varphi_i-\psi)\ .
\label{HesT}
\end{equation}
Neglecting the off-diagonal contributions
(their contribution changes the sign of at most one of the $N$ eigenvalues \cite{Angelani03})
the eigenvalues $\lambda_k$ of the Hessian calculated at the stationary point $\overline{\varphi}$
are obtained substituting equation (\ref{stazp}) in (\ref{HesT}),
\begin{equation}
\lambda_k=(-1)^{m_k}4 \xi^3\ .
\end{equation}
Therefore, the stationary point order $\nu(\overline{\varphi})$, defined as the number of negative eigenvalues
of the Hessian matrix, is simply the number of $m_k=1$ in the set \textbf{m} associated with $\overline{\varphi}$;
we can identify the quantity $n(\overline{\varphi})$ given by equation (\ref{ndef}) with the fractional order
$\nu(\overline{\varphi})/N<1/2$ of $\overline{\varphi}$.
Then, from equation (\ref{Ham}) and (\ref{xin}) 
we get a relation between the fractional order
$n(\overline{\varphi})$ and the potential energy $e(\overline{\varphi})=H(\overline{\varphi})/N$
at each stationary point $\overline{\varphi}$. It reads
\begin{equation}
n=\frac{1}{2}\left[1-(1-e)^{1/4}\right]\ ,
\label{n}
\end{equation}
where we have used the condition $n<1/2$.\\
Equation (\ref{n}) brings the condition
\begin{equation}
1-e>0,
\end{equation}
so there are no stationary points for $e>1$, while for $e<1$ the fractional order $n=\nu/N$
of the stationary points is a well defined monotonic function of their potential energy $e$,
given by equation (\ref{n}).\\
The number of stationary points of a given order
(apart a degeneracy factor)
is proportional to the number of ways in which one
can choose $\nu$ times 1 among the $\{m_k\}$, i.e. $\binom{N}{\nu}$.
Following \cite{Angelani03}, its logarithm 
\begin{equation}
\begin{array}{l}
\sigma(e)=\lim_{N\to\infty}\frac{1}{N}\ln\binom{N}{Nn(e)}=\\-n(e)\ln n(e)-(1-n(e))\ln (1-n(e)) \ ,
\end{array}
\end{equation}
represents the \textit{configurational entropy} of the saddles.
Substituting in this expression equation (\ref{n}) we have
\begin{align*}
\sigma(e)=-\frac{1}{2}\left[1-(1-e)^{1/4}\right]\ln \left[\frac{1}{2} \left[1-(1-e)^{1/4}\right]\right]-
\end{align*}
\begin{equation}
- \frac{1}{2}\left[1+(1-e)^{1/4}\right]\ln \left[\frac{1}{2}\left[1+(1-e)^{1/4}\right]\right] \ .
\end{equation}
For $e>1$ indeed we have, obviously, $\sigma(e)=0$.
This quantity is related to the Euler characteristic $\chi$ of the manifolds
$M_e=\{\varphi|H(\varphi)\leq N e\}$ \cite{Angelani03}
and its singular behavior around
the point $e=1$ is related to both the presence and the order
of the phase transition that occurs.\\
In figure \ref{fig3} the quantity $\sigma$ is reported as a function of energy $e$:
one can see that the presence
of a phase transition is signaled by a singularity
of $\sigma(e)$ at the transition point $e=1$.
It is worth noting that the curvature of the quantity $\sigma(e)$ around the transition point $e=1$
is positive, according to what found in Ref. \cite{Angelani03} for first order transitions.

Summarizing, the study of stationary points shows that the presence of the phase transition is signaled by the stationary point properties.
More specifically, the singular behavior of the configurational entropy $\sigma (e)$ at
transition point $e=1$ is the topological counterpart of the thermodynamic transition.
We note that these findings do not allow to discriminate between the {\it strong} and {\it weak}
topological hypothesis, as in this case the map $M$ from equilibrium energy levels to stationary
point energies is trivially the identity at the transition point  $e=1$: $M(1)=1$ \cite{Zamponi03}.

\section{Coherence properties and dynamics}
\label{secDyn}
The dynamics of interacting lasing modes close to the mode-locking transition can also
be investigated; it leads to explicit results for measurable
correlation functions.
\subsection{Single-mode first order coherence}
We start considering the single-mode (``self'') first order coherence \cite{LoudonBook}:
\beq
g^{(1)}_n (t)=\frac{\langle E^*_n(t_0) E_n(t_0+t)\rangle}{\langle E^*_n(t_0) E_n(t_0)\rangle}\ ,
\eeq
where $E_n(t)=\sqrt{\omega_n} a_n \exp(-i\omega_n t)$ is the electric field emitted at 
the angular frequency $\omega_n$ (omitting an inessential factor depending on the
point in space where the field is measured). 
For quenched amplitudes, $\omega_n A_n^2\cong \omega_0 A^2$, one has
(omitting the mode-index $n$)
\beq
g^{(1)} (t)=\frac{F^{(1)} (t)}{F^{(1)}(0)} e^{-i\omega_0 t} \ ,
\eeq
where $F^{(1)}(t)$ is the unnormalized single mode first order coherence given by
\beq
\label{F1n}
F^{(1)}(t)=\langle e^{i\varphi(t)} e^{-i\varphi(0)}\rangle \ ,
\eeq
and we used the fact that at equilibrium the time average over $t_0$ 
can be replaced by a statistical average
\cite{LoudonBook}.
Below the threshold ($T>T_c$) the phases are uniformly distribute in $[0,2\pi)$,
 so that $F^{(1)}(t\rightarrow\infty)=0$. 
In contrast, at mode-locking the phase is blocked around a fixed value, 
thus $F^{(1)}(t\rightarrow\infty)\neq 0$.

We are interested in the time-delay-profile of coherence function given by
\beq
F(t)=F^{(1)}(t)-F^{(1)}(\infty) \ .
\eeq
$F(t)$ can be explicitly calculated in the mean field theory  
(see Ref.~\cite{Zamponi03} for all the details of the computation).
In fact it can be shown that the single mode dynamics can be mapped into the
effective equation
\beq\label{dynabase}
\gamma \dot \varphi(t) = - 4 \zeta^3 \sin \varphi(t) + \eta(t) \ ,
\eeq 
where $\zeta$ is the thermodynamic value of the magnetization we determined
in section (\ref{secThe}), $\eta$ a $\delta$-correlated Gaussian noise with variance 
$2\gamma T$, and   $\gamma$ a constant fixing the time-scale 
(in the following we use units such that $\gamma=1$).

\begin{figure}[t]
\begin{center}
\includegraphics[width=0.45\textwidth]{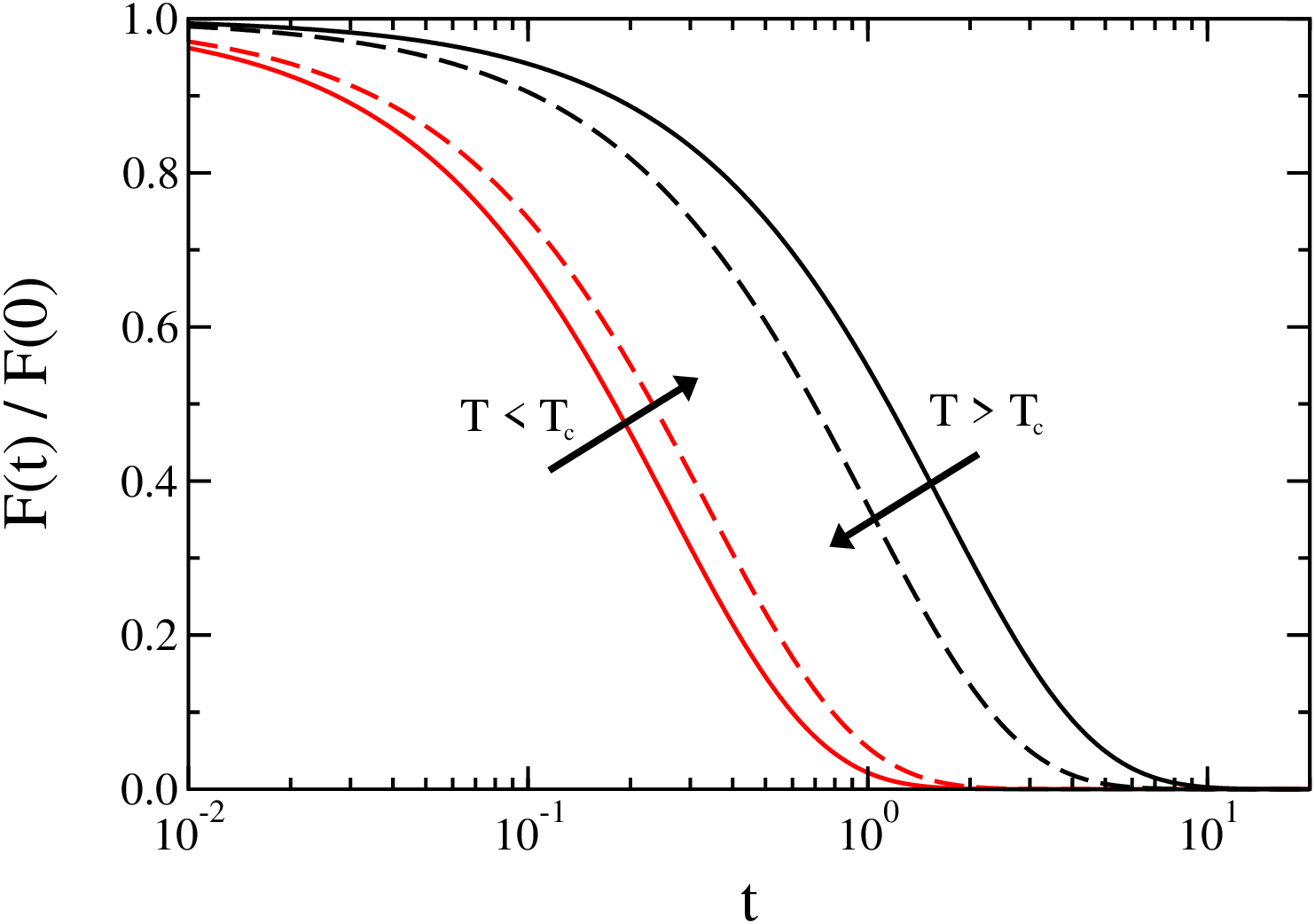}
\end{center}
\caption{(Color online)
Time dependence of the self correlation function $F(t)$ at different temperatures.
From left to right: $T=0.1, 0.5, 0.6, 1.0$ ($T_c=0.548$).
Time is in normalized units.}
\label{fig4}
\end{figure}

From Eq.~(\ref{dynabase}) it is evident that in the
paramagnetic phase ($\z=0$) the phases will freely diffuse, while in the ordered
(mode-locked) phase they fluctuate around a given value that, without loss of generality, can 
be taken as $\varphi =0$.
Eq.~(\ref{dynabase}) can be solved and the self-correlation function of a single mode
\beq
F(t) = \langle e^{i\varphi(t)} e^{-i\varphi(0)}\rangle - 
\langle e^{i\varphi(t)} \rangle \langle e^{-i\varphi(0)} \rangle 
\eeq
can be computed \cite{Zamponi03}.
Using symmetry properties, the above function can be written as
\beq
F(t) = F_c(t) + F_s(t) \ ,
\eeq
where 
\beq\label{fcs}
\begin{split}
F_c(t) &= \langle \cos\varphi(t) \cos\varphi(0) \rangle - 
\langle \cos \varphi(t) \rangle \langle \cos \varphi(0) \rangle\\
F_s(t) &= \langle \sin\varphi(t) \sin\varphi(0) \rangle \ .
\end{split}
\eeq
Following \cite{Zamponi03} the self correlations (\ref{fcs}) can be numerically determined
and they turn out to be nearly exponential at all temperatures,
$F_c(t) \propto e^{-t/\t_c}$ and $F_s(t) \propto e^{-t/\t_s}$.
In Fig.\ref{fig4}, the function $F(t)/F(0)$ is shown for different temperatures.
Upon increasing $T$ the decorrelation time increases for $T<T_c$,
while it decreases for $T>T_c$ (after a sudden jump at $T_c$).
This behavior is evident analyzing the $T$-dependence of the relaxation time.
In the upper panel of Fig.\ref{fig5} the quantity $\t_c$ 
is shown as a function of temperature. 
Full lines refer to stable states, while dashed lines to unstable ones.
We note that in paramagnetic high-$T$ phase $\t_c$ (and $\t_s$) have a $1/T$-dependence,
as expected for free Brownian motion.
In the low-$T$ phase the behavior of $\t_s$ (not showed in the Figure) is very similar to 
that of $\t_c$. \\
Summarizing, in absence of mode-locking the single-mode first-order coherence function
has an exponential trend (corresponding to a Lorentzian linewidth), whose relaxation time
decreases as the average energy per mode is reduced (i.e. the temperature is increased).
At the mode-locking transition, the coherence function is expressed as the sum of two exponentials
(corresponding to the two quadratures of the phase-modulated laser signal), whose time-constants
have a jump with respect to the ``free-run'' regime, and decreases while increasing the 
average energy per mode (and hence reducing the temperature).

\subsection{Multi-mode first order coherence}

Here we consider the multi-mode (``collective'') first order coherence:
\beq\label{cohe1def}
g^{(1)}(t)=\frac{\langle E^*(t_0)E(t_0+t)\rangle}{\langle E^*(t_0)E(t_0)\rangle}\text{,}
\eeq
with $E(t)=\sum_n \sqrt{\omega_n} a_n \exp(-i \omega_n t)$. 
In the quenched amplitudes approximation,
proceeding as above, it is possible to write
\beq
g^{(1)}(t)=\frac{G^{(1)}(t)}{G^{(1)}(0)}e^{-i\omega_0 t}\text{,}
\eeq
where we have taken $\omega_n\cong \omega_0$ (since all the modes are taken as densely packed
around $\omega_0$, small differences between $\omega_n$ and $\omega_0$ can be embedded in the
phase $\varphi_n$), and
\beq
G^{(1)}(t) = \frac{1}{N^2} \sum_{nm} \langle e^{i \varphi_n(t)} e^{-i \varphi_m(0)} \rangle =
\langle z(t) z^*(0) \rangle  \ ,
\eeq 
is the correlation function of the magnetization 
$z(t)=N^{-1} \sum_n \exp{[i \varphi_n(t)]}$.
\begin{figure}[t]
\begin{center}
\includegraphics[width=0.45\textwidth]{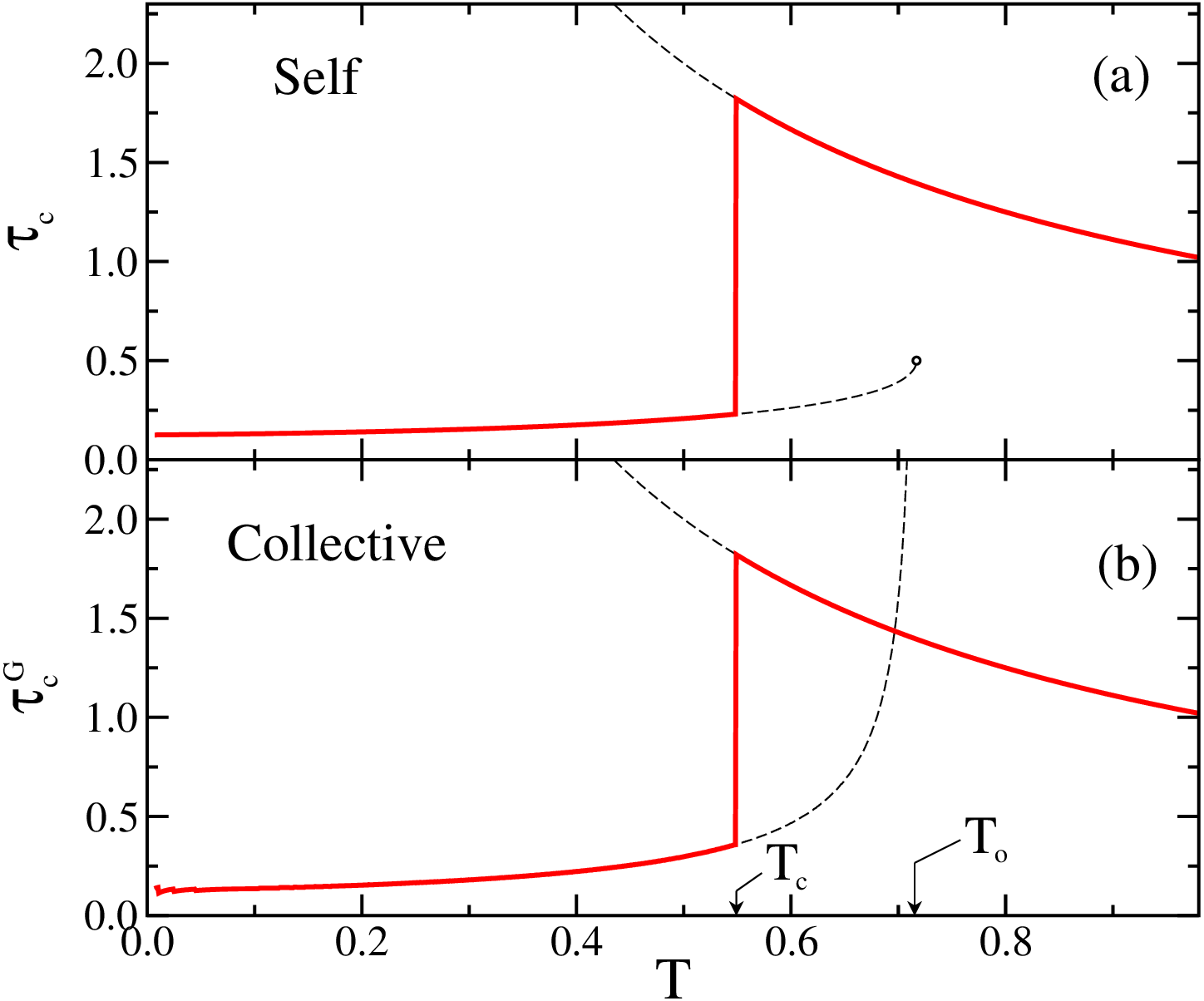}
\end{center}
\caption{(Color online)
(a) Relaxation time $\t_c$ of the self correlation function $F_c(t)$ as a function of $T$.
Full line corresponds to stable state (paramagnetic above $T_c$ and ferromagnetic below it).
Dashed lines refer to unstable solutions.
(b) Relaxation time $\t^G_c$ of the collective correlation function $G_c(t)$ as a function of $T$.
Full and dashed lines as before.
}
\label{fig5}
\end{figure}
We can write
\beq\label{G1G}
G^{(1)}(t) = G^{(1)}(\infty) + \frac1N G(t) = \z^2 + \frac1N G(t) \ ,
\eeq
where $G^{(1)}(\infty) =  \z^2$ is the asymptotic value 
(we recall that $\z = \langle z \rangle$ is assumed real), 
which is acquired at the mode-locking transition, and
the collective {\it connected} correlation function $G(t)$ is defined as
\beq\label{collective}
\begin{array}{l}
G(t) = N[ G^{(1)}(t)-G^{(1)}(\infty)]=\\
\displaystyle \frac{1}{N} \sum_{nm} [
\langle e^{i \varphi_n(t)} e^{-i \varphi_m(0)} \rangle - 
\langle e^{i \varphi_n(t)} \rangle \langle e^{-i \varphi_m(0)} \rangle ]  \ .
\end{array}
\eeq
This function has a finite limit for $N \to \io$, see the Appendix, which
can be computed following \cite{Zamponi03}. 
As for $F(t)$, $G(t)$ can be written as a sum of two terms
\beq
G(t) = G_c(t) + G_s(t) \ ,
\eeq
with
\beq\label{gcs}
\begin{split}
G_c(t) &= \frac{1}{N} \sum_{nm} [ \langle \cos\varphi_n(t) \cos\varphi_m(0) \rangle  - \\
& -  \langle \cos\varphi_n(t) \rangle \langle \cos\varphi_m(0) \rangle ] \ , \\
G_s(t) &= \frac{1}{N} \sum_{nm} [ \langle \sin\varphi_n(t) \sin\varphi_m(0) \rangle \ .
\end{split}
\eeq
Again one finds an exponential decay, 
$G_c(t) \propto e^{-t/\t^G_c}$ and $G_s(t) \propto e^{-t/\t^G_s}$  \cite{Zamponi03}.
In the lower panel of Fig.\ref{fig5} the relaxation time $\t^G_c$ 
is shown as a function of temperature. 
We hence expect a trend for $G(t)$ which resembles $F(t)$; however
it is worth noting that the quantity $\t^G_c$ diverges
when $T \rightarrow T_o^-$, that is, when, starting from low $T$ phase, 
the point where the unstable solution (dashed line) disappear is approached. 
The occurrence of the thermodynamic transition at $T_c < T_o$ prevents the 
divergence of $\t^G_c$.  The $\t^G_s$ (not shown) does not diverge at $T_o$.\\

\subsection{Multi-mode second order coherence}

We also consider the multi-mode (collective) second-order coherence:
\beq\begin{split}
g^{(2)}(t)&=\frac{\langle E^*(t_0) E^*(t_0+t) E(t_0) E(t_0+t)\rangle}
{\langle E^*(t_0) E(t_0)\rangle^2} \\ 
&=\frac{\langle I(t_0) I(t_0+t)\rangle}
{\langle I(t_0)\rangle^2} \ .
\end{split}
\eeq
again with 
\beq
E(t)=\sum_n \sqrt{\omega_n} a_n e^{-i \omega_n t}
\propto z(t) e^{-i \omega_0 t} \ .
\eeq
We get
\beq
g^{(2)}(t) = \frac{G^{(2)}(t)}{G^{(2)}(\io)} \ ,
\eeq
where
$G^{(2)}(t) = \langle z(t) z^*(t)z(0) z^*(0) \rangle$.
We can decompose this function in its connected
components (see Ref.\cite{ParisiBook}, Eq. 4.23); 
recalling that $\z = \langle z \rangle$ is
assumed real we get
\beq\begin{split}
G^{(2)}(t) &= \langle z(t) z^*(t)z(0) z^*(0) \rangle_c +
\z G_3(t) \\
&+ \langle z(0) z^*(0) \rangle_c^2 +
| \langle z(t)  z(0) \rangle_c|^2  \\
&+ | \langle z(t)  z^*(0) \rangle_c|^2
+ \z^4
\ ,
\end{split}
\eeq
where the function $G_3(t)$ is a sum of
connected three-point functions:
$ G_3(t) = 2 \re [
\langle z(t) z^*(t) z(0) \rangle_c + 
 \langle z(t) z(0) z^*(0) \rangle_c ]$.
Using the results of the Appendix,
we have
$\langle z(t) z^*(t)z(0) z^*(0) \rangle_c \propto N^{-3}$
and $G_3 \propto N^{-2}$;
from Eq.s~(\ref{collective}), (\ref{gcs}), we have
\beq\begin{split}
&\langle z(0) z^*(0) \rangle_c = N^{-1} G(0) \ , \\
&\langle z(t) z(0) \rangle_c = N^{-1} [G_c(t)-G_s(t)] \ , \\
&\langle z(t) z^*(0) \rangle_c = N^{-1} [G_c(t)+G_s(t)] \ .
\end{split}
\eeq
Then we obtain
\beq\label{G2finale}
\begin{split}
G^{(2)}(t) &= \z^4 + \frac{G(0)^2 + 2 G_c(t)^2 + 2 G_s(t)^2 }{N^2} \\
&+ \z G_3(t) + O(N^{-3}) \ .
\end{split}
\eeq

In the paramagnetic phase, $\z=0$, and by symmetry $G_s(t) = G_c(t) = G(t)/2$;
moreover these functions tend to $0$ for $t\to\io$.
Then we get
\beq\label{G2G1}
\begin{split}
g^{(2)}(t) &= \frac{G(0)^2 + 2 G_c(t)^2 + 2 G_s(t)^2}{G(0)^2} \\ &=
1 + \frac{G(t)^2}{G(0)^2} = 1 + |g^{(1)}(t)|^2 \ ,
\end{split}
\eeq
where we make use also of Eq.~(\ref{G1G}). This result is indeed what
we expect for light modes evolving independently and rapidly. 

In the mode-locked phase Eq.~(\ref{G2G1}) will not hold but 
a relation between $g^{(2)}$ and $g^{(1)}$
can still in principle be deduced from the knowledge of 
function $G_3(t)$, 
using Eq.~(\ref{G2finale}) and
(\ref{G1G}).  
Future works will address this point.

\section{Conclusions}
\label{secCon}
By using a simple model that is expected to describe multi-mode dynamics of
tightly packed extended and/or localized modes in a nano-optical resonator,
we predict the existence of a first-order phase-locking transition
when the averaged energy per mode is above a critical value
(correspondingly the adimensional effective temperature is below $T_c$).
This value depends on the average value of the mode-overlap coefficient $g$.
If the transition involve extended modes, one has (omitting indexes)
$g\cong \omega_0^2 \int |E|^4 dV \cong \chi_0 \omega_0^2 V^{-1}$,
with $\chi_0$ a reference susceptibility value.
Conversely if localized modes are involved it is $g\cong \chi_0 V_0^{.2}\omega_0^2 $
where $V_0$ is the average localized mode volume ($V_0\propto L_0^2$ with
$L_0$ the localization length).
In the former case,
for a fixed spontaneous emission noise $T_{bath}$, it is found that the
critical mean energy per mode is $V$-dependent
\begin{equation}
\mathcal{E}^{(ext)}_c=\omega_0 A^2= \sqrt{\frac{k_B T_{bath} V}{T_c \chi_0}} \ ,
\end{equation}
while for localized modes
\begin{equation}
\mathcal{E}^{(loc)}_c=\omega_0 A^2= \sqrt{\frac{k_B T_{bath} L_0^3}{T_c \chi_0}} \ .
\end{equation}
Hence the critical energy
for the phase-locking transition has very different scaling behavior with respect to the system size, depending
on the degree of localization of the involved modes.
In the general case one can expect intermediate regimes between those considered, so that
the trend of the critical energy (determined by the amount of energy pumped in the system per unit time, i.e. the pumping rate)
versus system volume is an interesting quantity  which can experimentally investigated.
A topology-thermodynamics relationship has been evidenced for our model,
corroborating previous findings on this topic: the thermodynamic transition is
signaled by a singularity in the topological quantity $\sigma$.

The exact solution of the dynamics of the model predicts the 
divergence of the relaxation time of the first-order coherence function $g^{(1)}$ at the transition. 
This behavior might be observed in experiments; the different 
scaling with respect to the systems size also affect the position of the transition, as determined
by the predicted jump in the relaxation time or, equivalently, in an abrupt change of the single-mode laser linewidth
while varying the pumping rate. 

This analysis points out the rich
phase-space structure displayed by these systems, 
while varying the amount of disorder or the profile
of the density of states. Hence nano-lasers not only may furnish the basis for
highly integrated short-pulse generators, but are
a valuable framework for fundamental physical
studies. These deserve future theoretical and experimental investigations and can
be extended to other nonlinear multi-mode interactions.

\section*{Appendix}

We discuss here the scaling of the correlations of $z$ when $N \to \io$.\\
The basic fact is that the variable $z$ is intensive and, 
for a mean field system, has a probability distribution
of the form~\cite{ParisiBook}
\beq
P_N(z) = e^{N F(z)} \ ,
\eeq
where $F$ is related to the thermodynamic free energy of the system.
Consider the generating functional
\beq
e^{N F(j)} = \langle e^{N j z} \rangle = \int dz e^{N [F(z)+jz]} \ ,
\eeq
then for large $N$, $F(j) = \max_{z} [F(z) + jz]$ and it is a quantity
of order 1.
The {\it connected correlation functions}~\cite{ParisiBook} 
are derivatives of $N F(j)$ with respect to $N j$:
\beq
\langle z^k \rangle_c = \left. \frac{\d^k N F(j)}{(\d N j)^k} \right|_{j=0} =
N^{1-k}  \left. \frac{\d^k F(j)}{(\d j)^k} \right|_{j=0} \ .
\eeq
This simple argument shows that $\langle z^k \rangle_c \propto N^{1-k}$.
In particular,
\begin{eqnarray}
\langle z^2 \rangle_c &=& \langle z^2 \rangle -\langle z \rangle^2 \propto N^{-1} \ , \nonumber\\
\langle z^3 \rangle_c &=& \langle z^3 \rangle - 3 \langle z \rangle
\langle z^2 \rangle_c - \langle z \rangle^3 \propto N^{-2} \ , \\
\langle z^4 \rangle_c &=& \langle z^4 \rangle - 4 \langle z \rangle
\langle z^3 \rangle_c - 3 \langle z^2 \rangle_c^2 -  \langle z \rangle^4 \propto N^{-3}\nonumber \ .
\end{eqnarray}
We assumed that $z$ is real but the same derivation can be repeated for
a complex variable; the only difference is in the definition of the connected
correlation functions.

For the dynamics, one can write a similar expression for the probability
of a trajectory $z(t)$, see \cite{Zamponi03} and references therein:
\beq
P_N[z(t)] = e^{N F[z(t)]} \ .
\eeq
Repeating the derivation above using functional integrals one obtains
exactly the same results for the scaling with $N$.


\end{document}